\documentclass[12pt]{article}
\usepackage[dvips]{graphicx}

\usepackage{graphics,epsfig}

\renewcommand{\bar}[1]{\overline{#1}}
\renewcommand{\bar}[1]{\overline{#1}}

\def\ru1{\rule[-0.4truecm]{0mm}{1truecm}}

\begin{document}

%\begin{titlepage}
\rightline{\small KCL-PH-TH/2011-27, LCTS/2011-13, CERN-PH-TH/2011-212}
\vspace{10mm}

\centerline{\Large \bf Spin Correlations of $\Lambda{\bar \Lambda}$ Pairs}
\vspace{3mm}
\centerline{\Large \bf as a Probe of Quark-Antiquark Pair Production}

\vspace{7mm}

\centerline{{\bf John Ellis$^{1}$} and {\bf Dae Sung Hwang$^{2}$}}

\vspace{5mm}

\vspace{4mm} \centerline{\it $^1$Theoretical Particle Physics and Cosmology Group, Physics Department,}
\centerline{\it King's College London, Strand, London, UK;}
\centerline{\it Theory Division, Physics Department, CERN, 1211 Geneva 23, Switzerland}

\vspace{4mm} \centerline{\it $^2$Department of Physics, Sejong
University, Seoul 143--747, South Korea}

\vspace{20mm}

\centerline{\bf Abstract}

\vspace{10mm}

\noindent
The polarizations of $\Lambda$ and ${\bar \Lambda}$ are 
thought to retain memories of the spins of their parent $s$ quarks and ${\bar s}$
antiquarks, and are readily measurable {\it via}
the angular distributions of their daughter protons and antiprotons. 
Correlations between the spins of $\Lambda$ and ${\bar \Lambda}$
produced at low relative momenta may therefore be
used to probe the spin states of $s {\bar s}$ pairs produced during hadronization.
We consider the possibilities that they are produced in a $^3$P$_0$ state, as might
result from fluctuations in the magnitude of $\langle {\bar s} s \rangle$, a $^1$S$_0$ state, 
as might result from chiral fluctuations, or a $^3$S$_1$ or other spin state, as might result from
production by a quark-antiquark or gluon pair. We provide templates for the $p {\bar p}$ angular correlations
that would be expected in each of these cases, and discuss how they might be used to
distinguish $s {\bar s}$ production mechanisms in $pp$ and heavy-ion collisions.

\vspace{0.5cm}

\noindent \vspace*{12mm}

%\noindent {\bf Keywords:}

\vspace{3mm}


%\end{titlepage}

%\setlength{\baselineskip}{13pt}

%\baselineskip 22pt

\newpage

%\vspace{1.0cm}

\section{Introduction}

Hadronization proceeds {\it via} the production of ${\bar q} q$ pairs, that may arise
via a combinations of perturbative and non-perturbative mechanisms,
such as gluon splitting $g \to {\bar q} q$ and fluctuations in the chiral
condensate $\langle {\bar q} q \rangle$. It is quite possible that the relative
importances of these mechanisms may depend on the types of particles
colliding, e.g., $pp$ or heavy-ion collisions, and on the kinematic conditions,
e.g., low momenta in minimum-bias events or at high $p_T$ inside jets.

These mechanisms suggest various different possibilities for the ${\bar q} q$
quantum numbers, and in particular their possible spin states. However,
it is not immediately apparent how one could determine these spin states 
by penetrating the `hadronization firewall' {\it via} measurements of final-state hadrons.
However, one tool for measuring quark spins indirectly is known, namely measuring
the polarization states of unstable final-state hyperons, particularly $\Lambda$ baryons~\cite{Lambda}. These may be
determined by measuring the angular distributions of their decay products, which are
in general of the form $(1 + P \alpha \cos \theta)$, where $P$ is the polarization and
$\alpha \sim 0.6$ in the case of $\Lambda \to p \pi^-$ decay~\cite{PDG}. Models of baryon spins
based on SU(6) wave functions suggest that the $\Lambda$ `remembers' very well
the polarization of its parent $s$ quark, with the accompanying $u d$ pair expected
to be in a spin-singlet state~\cite{models}. Experimentally, this naive picture seems to be
qualitatively correct, e.g., from measurements of $\Lambda$ polarization in final
states resulting from $s$ quarks with known spin states~\cite{expts}.

Here we go one step further by proposing to use measurements of the angular
correlations between the ${\bar p}$ and $p$ produced in the decays of ${\bar \Lambda} \Lambda$
pairs to analyze the spin states of parent ${\bar s} s$ pairs, specifically those with small
relative momenta that could have been produced by a common production reaction.

In the case of perturbative $g \to {\bar s} s$ splitting, the final state pair would be in a
vector state, that could correspond to  a $^3$S$_1$ or $^3$D$_1$ configuration. The former
would dominate if the strange quark mass could be neglected, but the latter is potentially
also important for massive quarks, as evidenced by the appearance of a $^3$D$_1$
${\bar c} c$ vector meson in $e^+ e^-$ annihilation. Both these configurations are
spin-triplet states, so in both cases one might expect the ${\bar \Lambda} \Lambda$ pair also to
have a spin-triplet configuration. However, whereas in the $^3$S$_1$ case the ${\bar \Lambda} \Lambda$
pairs could be expected to have parallel polarizations, this is not necessarily
the case in the $^3$D$_1$ case. In the case of perturbative $gg \to {\bar s} s$
production with centre-of-mass energy $\sqrt{s}$, 
other configurations for the ${\bar \Lambda} \Lambda$ spin correlations become possible,
interpolating between $^3$S$_1$ if the $s$ quark mass can be neglected to a spin-singlet
configuration if $\sqrt{s} = 2 m_s$.

In the non-perturbative case, models for ${\bar s} s$ and ${\bar \Lambda} \Lambda$ pair 
production would take their inspiration from our understanding of chiral dynamics.
In the standard QCD vacuum, it is known that $\langle 0 | {\bar q} q | 0 \rangle \ne 0$
for $ q = u, d, s$~\cite{GOR}, and the lowest-lying pseudoscalar mesons correspond to chiral spin waves~\cite{Nambu},
i.e., spatial fluctuations in the chiral orientation of the $\langle 0 | {\bar q} q | 0 \rangle$ condensate.
It is also known that at high temperatures, such as those that may be attained in
heavy-ion collisions, the quark condensates $\langle 0 | {\bar q} q | 0 \rangle \to 0$~\cite{lattice}, whereas
perturbative calculations of `hot' initial states assume implicitly that the
$\langle 0 | {\bar q} q | 0 \rangle$ can be neglected. Therefore, it seems possible that
either (i) the magnitude of $\langle {\bar q} q \rangle$ varies during the
hadronization process and/or (ii) that chiral spin `ripples' with ${\bar q} \gamma_5 q \ne 0$
are  generated during hadronization. The former might lead to production of
${\bar s} s$ pairs in a scalar $^3$P$_0$ configuration, and the latter to 
${\bar s} s$ pairs in a pseudoscalar $^1$S$_0$ configuration. In both cases, the
 ${\bar \Lambda}$ and $\Lambda$ spins would be in a totally anti-correlated spin-singlet state.

In order to select ${\bar \Lambda} \Lambda$ pairs that are most likely to be due to
pair-production of a single ${\bar s} s$ pair, we propose to examine
${\bar \Lambda} \Lambda$ pairs with small relative 3-momenta ${\mathbf p}$.
In the cases of S-wave configurations, namely the $^3$S$_1$ and $^1$S$_0$
mentioned above, there would be no correlation between the directions of ${\mathbf p}$
and the ${\bar \Lambda}$ and $\Lambda$ spins. In the P- and D-wave cases
$^3$P$_0$ and $^3$D$_1$, such a correlation could be expected, but we do not discuss this
possibility here.

In this paper we calculate these spin and momentum correlations for all the
possibilities discussed above and evaluate the possibility of measuring them
in the hadronic final states produced in $pp$ and/or heavy-ion collisions. We note
again that the dominant hadronization processes in these two classes of reactions
might be different. Specifically, the final states in heavy-ion collisions are thought to
have evolved from a thermal plasma, albeit a strongly-interacting one in which the
relevant degrees of freedom close to the phase transition might not be the
conventional perturbative quarks and gluons. The type of analysis proposed here
might provide some insight into the nature of the relevant degrees of freedom. On
the other hand, different mechanisms are likely to come into play in $pp$
collisions, which are unlikely to have been thermal and might be perturbative at
high ${\mathbf p}_T$. The type of analysis proposed here might provide
interesting insights into the similarities and/or differences between hadronization
mechanisms in $pp$ and heavy-ion collisions.

\section{$\Lambda{\bar \Lambda}$ Spin Correlation as a Discriminant between Models of ${\bar s}s$ Production}

The polarization of the $\Lambda$ (${\bar \Lambda}$) can be measured from the angular
distribution of the daughter particles in the decay channel $\Lambda \to p\pi^-$
(${\bar \Lambda} \to {\bar p}\pi^+$).
The angular distribution of the final-state (anti-)proton in the $\Lambda$ (${\bar \Lambda}$)
rest frame is given by
\begin{equation}
{dN\over d{\cos{\theta^*}}} \; = \; {N_{\rm tot}\over 2}(1+\alpha P{\cos{\theta^*}})\ ,
\label{cc1}
\end{equation}
where $N_{\rm tot}$ is the total number of $\Lambda$ (${\bar \Lambda}$),
$\alpha = +(-)0.642\pm 0.013$ is the $\Lambda$ (${\bar \Lambda}$) decay parameter~\cite{PDG},
$P$ is the $\Lambda$ (${\bar \Lambda}$) polarization, and $\theta^*$ is the angle between
the (anti-)proton momentum and the $\Lambda$ (${\bar \Lambda}$) polarization
direction in the $\Lambda$ (${\bar \Lambda}$) rest frame.
Corresponding to (\ref{cc1}), the double angular distribution for $\Lambda {\bar \Lambda}$
pair production with polarizations $P_{1,2}$ and centre-of-mass decay angles $\theta_{1,2}^*$ is given by 
\begin{equation}
{d^2 N\over d{\cos{\theta_1^*}} d{\cos{\theta_1^*}}} \; = \;
{N_{\rm tot}\over 4}(1+\alpha_1 P_1{\cos{\theta_1^*}})(1+\alpha_2 P_2{\cos{\theta_2^*}}) \ ,
\label{cc2}
\end{equation}
where $\alpha_2 = - \alpha_1$ for particle-antiparticle pairs,
and our next task is to estimate $P_{1,2}$ in different models for $\Lambda {\bar \Lambda}$
pair production.

\subsection{Production via a Scalar or Pseudoscalar Coupling}

\begin{figure}
\centering
%\psfrag{xperp}[cc][cc]{$x_\perp$}
%\vspace*{0.5cm}
%\begin{minipage}[t]{9.5cm}
\begin{minipage}[t]{9.5cm}
\centering
\includegraphics[width=\textwidth]{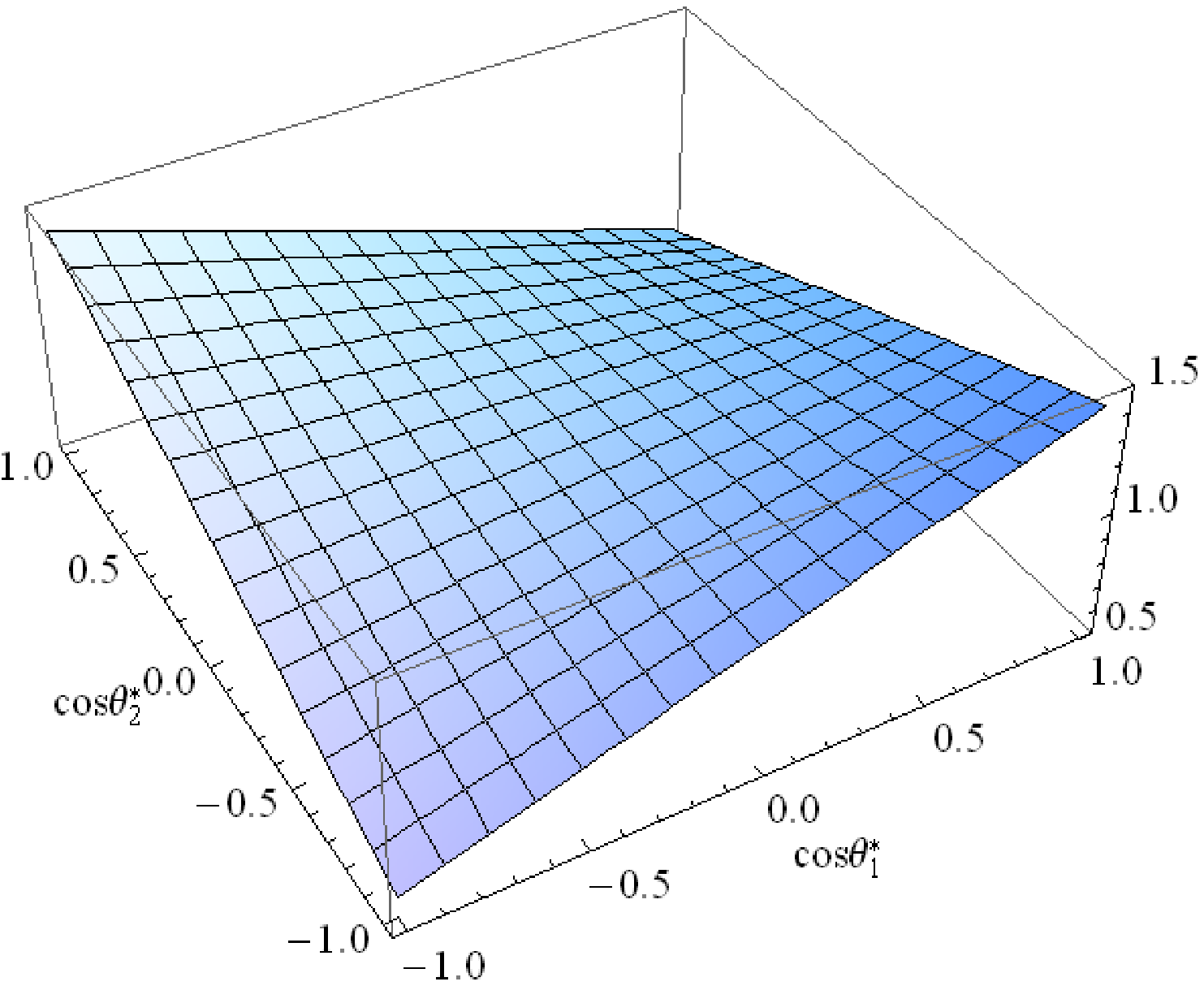}
\includegraphics[width=\textwidth]{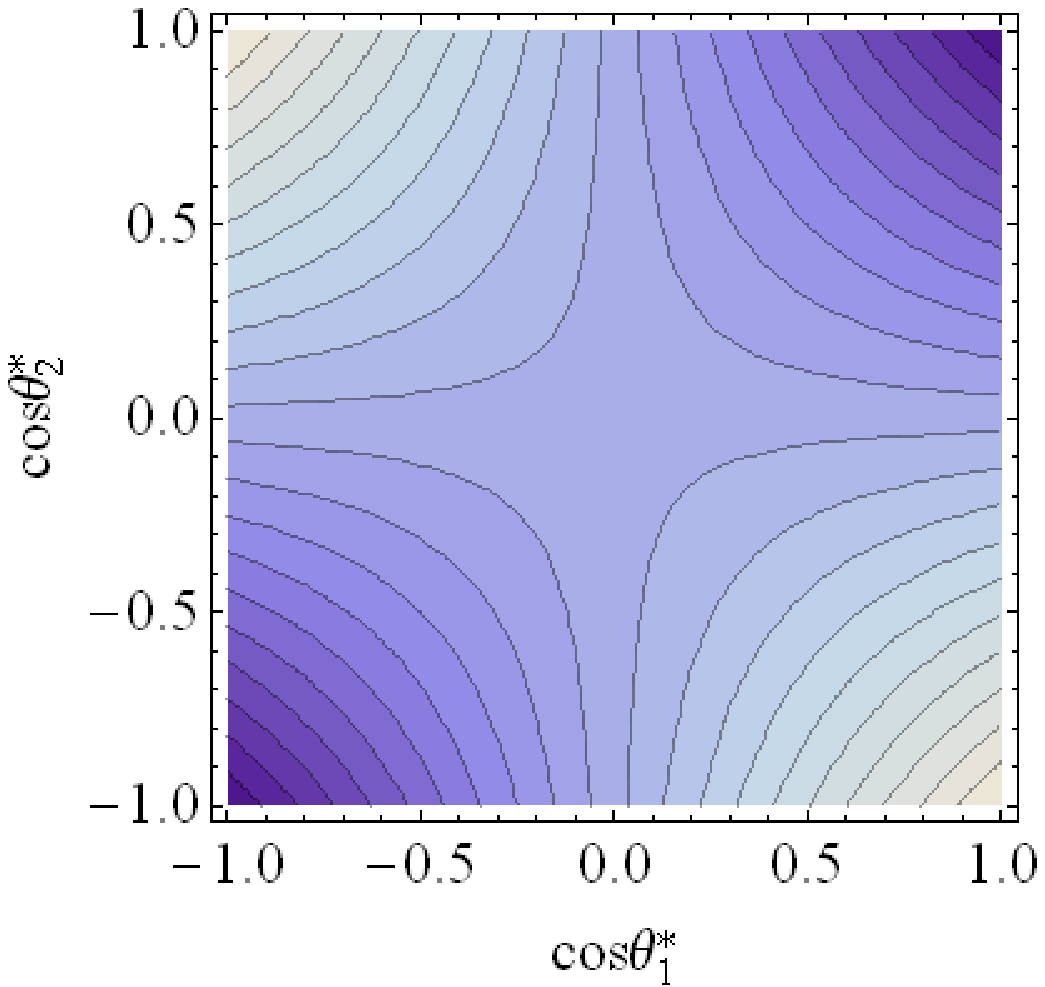}
%(a1)
\end{minipage}\hspace{0.0cm}
\vspace*{1.0cm}
%%%%
%%%%
\parbox{0.95\textwidth}{\caption{
\it Correlation between $\cos{{\theta}^*_1}$ and $\cos{{\theta}^*_2}$ for a scalar
or pseudo-scalar coupling.
\label{fig:SP}}}
\end{figure}

Production of ${\bar s}s$ pairs in a scalar $^3$P$_0$ or pseudoscalar $^1$S$_0$ state
might be favoured in some non-perturbative scenarios. In particular, as already commented
in the Introduction, the transition from the perturbative (or high-temperature) vacuum
with $\langle s {\bar s} \rangle = 0$ that
might be relevant at short distances (or high densities and pressures) to the non-perturbative
vacuum with $\langle s {\bar s} \rangle \ne 0$ relevant at large distances (or low temperatures)
maybe accompanied by fluctuations in the modulus of the ${\bar s}s$ condensate that could
manifest themselves as $^3$P$_0$ ${\bar s}s$ pairs. Alternatively, during this transition
there might arise chiral spin waves that could manifest themselves as $^1$S$_0$ ${\bar s}\gamma_5 s$ pairs.
In either case, the ${\bar s}s$ pair is produced in a spin-singlet state, and hence the
polarizations of ${\bar s}$ and $s$ would be either both along their momentum directions, or
both opposite to their momentum directions. Furthermore, the amplitudes for these two states
would have the same magnitudes, and they would not interfere. Hence we may add incoherently
contributions of the form (\ref{cc2}) with $P_1 = +1, P_2 = +1$ and with $P_1 = -1, P_2 = -1$,
obtaining a decay-angle correlation that is proportional to:
\begin{equation}
{1\over 2}\Big[ (1+a{\cos{\theta^*_1}})((1-a{\cos{\theta^*_2}})
+(1-a{\cos{\theta^*_1}})((1+a{\cos{\theta^*_2}})\Big] =
(1-a^2{\cos{\theta^*_1}}{\cos{\theta^*_2}})\ ,
\label{cc3}
\end{equation}
where $a=0.642\pm 0.013$ is the $\Lambda$ decay parameter~\cite{PDG}.
Fig.~\ref{fig:SP} displays the between $\cos{{\theta}^*_1}$ and $\cos{{\theta}^*_2}$ to be
expected on the basis of (\ref{cc3}) in the case of a scalar or pseudo-scalar coupling. 

\subsection{Production via a Vector Coupling}

As alternatives, we consider a couple of perturbative production mechanisms,
namely the process ${\bar q} q \to {\bar s} s$ that is mediated by gluon exchange and hence via a
vector coupling, or the process $gg \to {\bar s}s$ to which several perturbative diagrams
contribute leading to a more complicated spin structure. In this subsection we consider the
${\bar q} q \to {\bar s} s$ case, initially assuming that the $s$ mass can be neglected.

\begin{figure}
\centering
%\psfrag{xperp}[cc][cc]{$x_\perp$}
%\vspace*{0.5cm}
\begin{minipage}[t]{9.5cm}
\centering
\includegraphics[width=\textwidth]{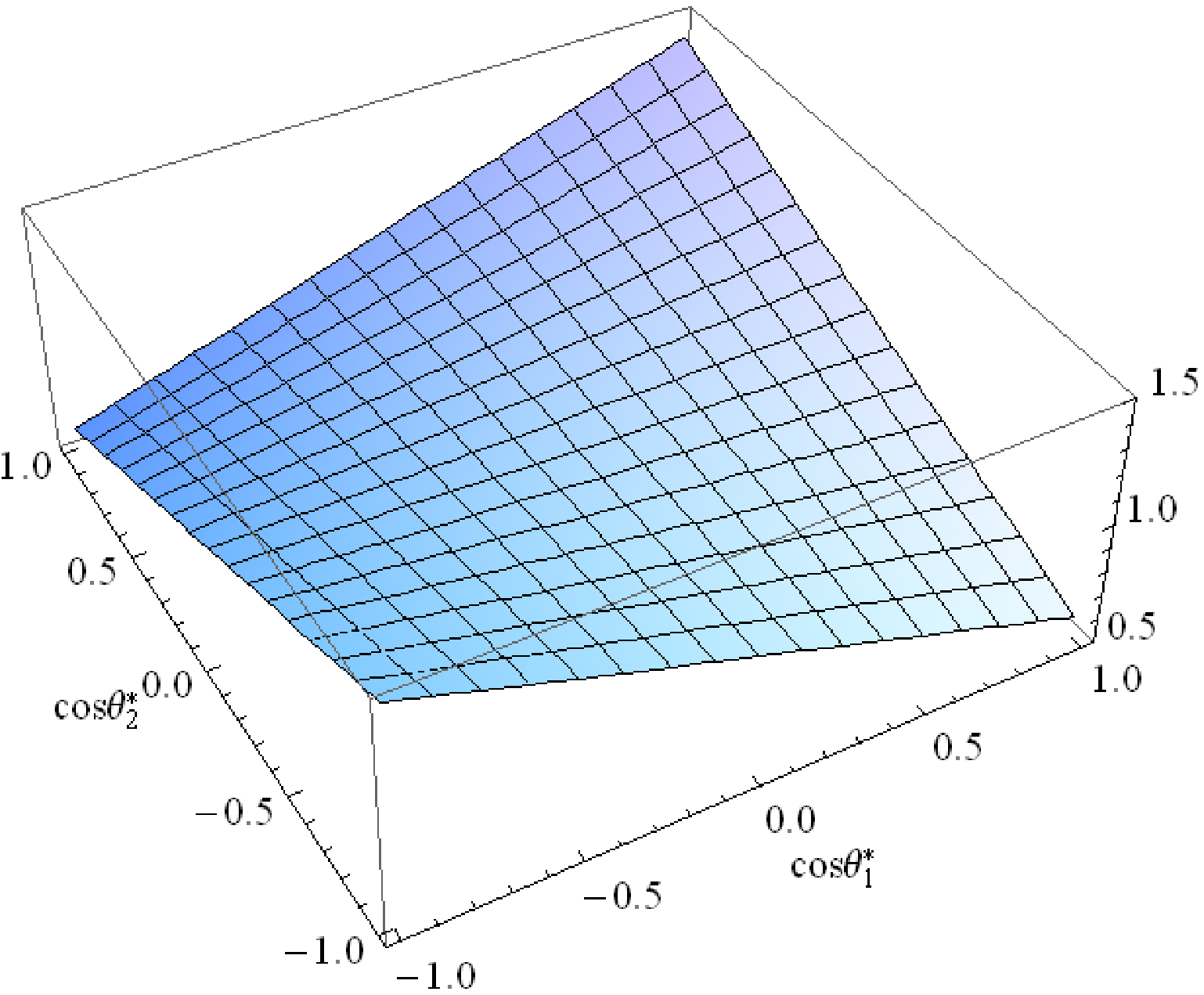}
\includegraphics[width=\textwidth]{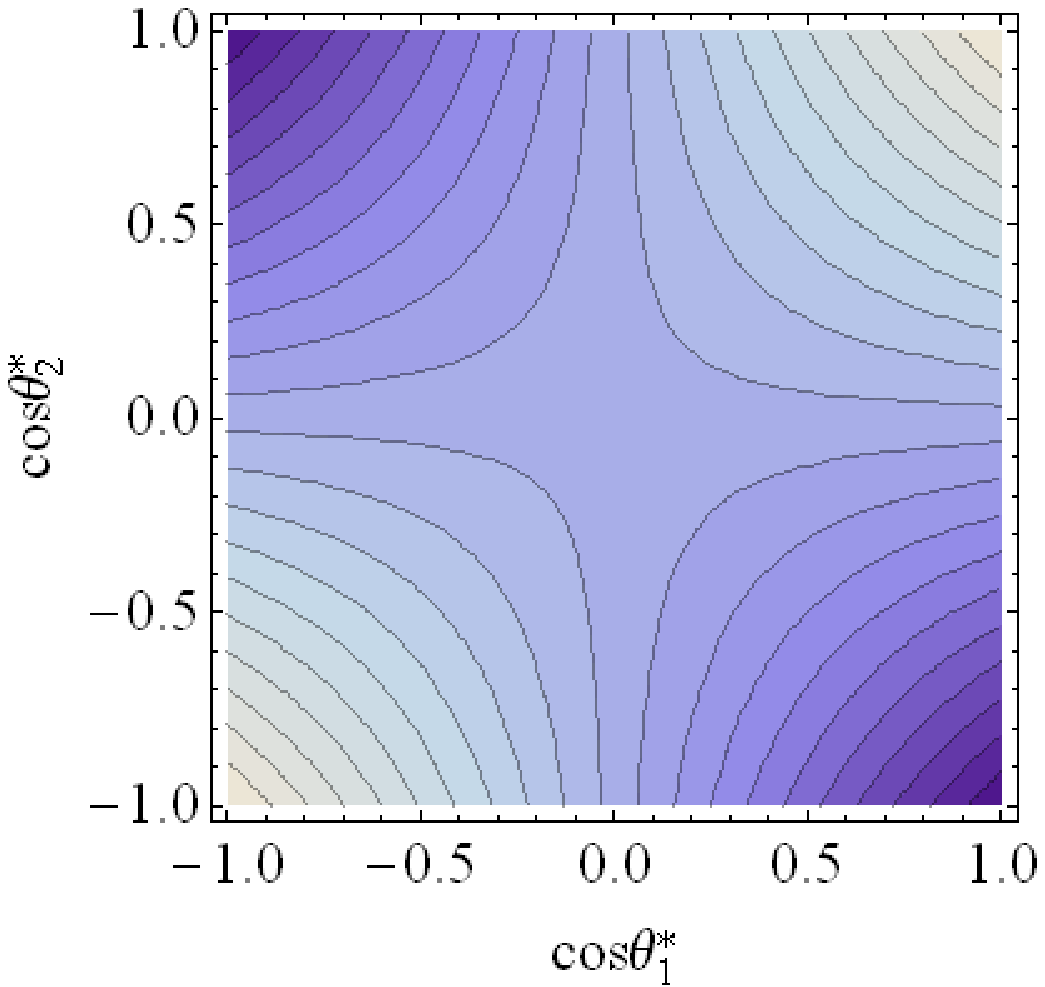}
%(a1)
\end{minipage}\hspace{0.0cm}
\vspace*{1.0cm}
%%%%
%%%%
\parbox{0.95\textwidth}{\caption{
\it Correlation between $\cos{{\theta}^*_1}$ and $\cos{{\theta}^*_2}$ for a vector coupling
when $m_s = 0$.
\label{fig:V0}}}
\end{figure}

In this case, there are only two combinations of the $s$ and ${\bar s}$ polarizations:
either the polarization of the $s$ is along and that of the ${\bar s}$ is opposite to its momentum direction
or the polarization of the $s$ is opposite and that of the ${\bar s}$ is along its momentum direction.
When we consider the collision of unpolarized proton and proton or unpolarized lead-lead
nuclei as in LHC, the cross sections for the above two combinations of
the $s$ and ${\bar s}$ polarizations are the same.
Hence we may add incoherently
contributions of the form (\ref{cc2}) with $P_1 = +1, P_2 = -1$ and with $P_1 = -1, P_2 = +1$,
obtaining a correlation that is proportional to:
\begin{equation}
{1\over 2}\Big[ (1+a{\cos{\theta^*_1}})((1+a{\cos{\theta^*_2}})
+(1-a{\cos{\theta^*_1}})((1-a{\cos{\theta^*_2}})\Big] =
(1+a^2{\cos{\theta^*_1}}{\cos{\theta^*_2}})\ ,
\label{cc4}
\end{equation}
where $a=0.642\pm 0.013$~\cite{PDG}.
Fig.~\ref{fig:V0} displays the between $\cos{{\theta}^*_1}$ and $\cos{{\theta}^*_2}$ to be
expected on the basis of (\ref{cc4}) in the case of a vector gluon coupling with $m_s = 0$. 
We see that this is, in principle, easily distinguishable from the scalar/pseudoscalar
case (\ref{cc3}), thanks to the completely different $\Lambda {\bar \Lambda}$
polarization correlations and the strong analyzing power of $\Lambda \to p \pi^-$
decay.

The case $m_s \ne 0$ is slightly more complicated, with a non-trivial combination
of $\Lambda {\bar \Lambda}$ polarization states becoming possible. An elementary
calculation of ${\bar q} q \to {\bar s} s$, averaging over the polarizations of the massless
quarks in the initial state and keeping track of the final-state polarizations, 
yields a decay-angle correlation that is proportional to:
\begin{equation}
\left[1+a^2{\cos{\theta^*_1}}{\cos{\theta^*_2}}\right] + \frac{x^2}{2}\left[1-a^2{\cos{\theta^*_1}}{\cos{\theta^*_2}}\right] \, ,
\label{cc5}
\end{equation}
where $a=0.642\pm 0.013$ is the $\Lambda$ decay parameter, as before, and
$x \equiv 2 m_s/\sqrt{s}$. This reduces to the case (\ref{cc4}) in the limit $m_s \to 0$,
but we see from the second term in (\ref{cc5}) that the spin correlation in the massless
case is diluted for $m_s \ne 0$, reflecting an admixture of the $^3$D$_1$ state. 
However, the spin correlation remains relatively large and of the
same sign for all masses. As a measure of this, we define a one-dimensional correlation
parameter $f(x)$ as follows:
\begin{equation}
{(TR+BL)-(TL+BR)\over (TR+BL)+(TL+BR)}\ \equiv\ {a^2\over 4}\ f(x)\ =\
{a^2\over 4}\
{2-x^2\over 2+x^2}\ ,
\label{dd7}
\end{equation}
where TR, BL, TL, and BR refer to the top-right, bottom-left, top-left and bottom-right quadrants,
respectively, in the lower panels of Figs.~\ref{fig:SP} and \ref{fig:V0}, i.e., T $\equiv \cos{\theta^*_2} > 0$,
B $\equiv \cos{\theta^*_2} < 0$, R $\equiv \cos{\theta^*_1} > 0$ and L $\equiv \cos{\theta^*_1} < 0$.

\begin{figure}
\centering
%\psfrag{xperp}[cc][cc]{$x_\perp$}
%\vspace*{0.5cm}
\begin{minipage}[t]{8.5cm}
\centering
\includegraphics[width=\textwidth]{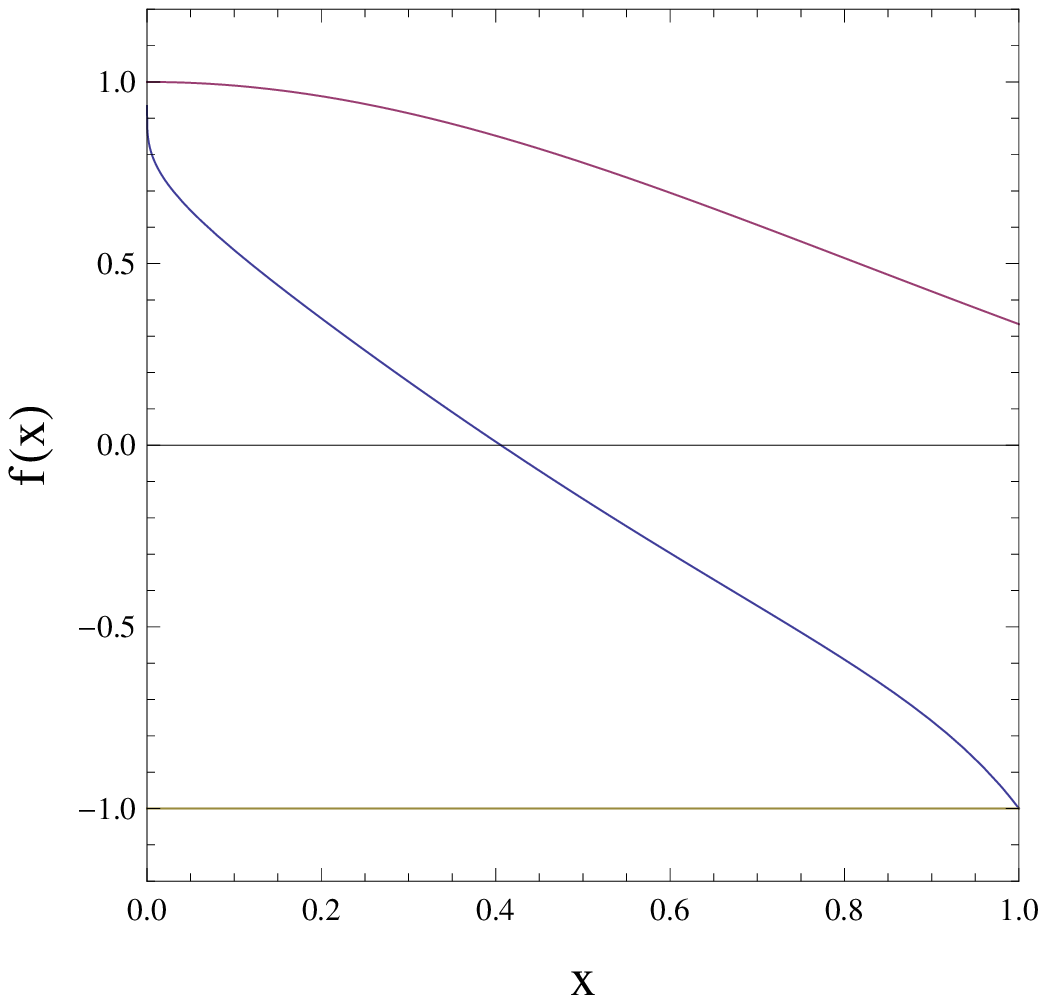}
%(a1)
\end{minipage}\hspace{0.0cm}
\vspace*{1.0cm}
%%%%
%%%%
\parbox{0.95\textwidth}{\caption{
%Correlation of $\cos{{\theta}^*_1}$ and $\cos{{\theta}^*_2}$ for
%the $q{\bar q}\to q{\bar q}\to \Lambda {\bar \Lambda}$ production mechanism.
\it Graphs of the correlation $f(x)$ defined in (\protect\ref{dd7})
as a function of $x\equiv {2m_s\over {\sqrt{s}}}$, where $m_s$ is the strange quark mass.
The topmost line is for the vector case (\protect\ref{cc5}), the lowest line is for the scalar and pseudoscalar
cases (\protect\ref{cc3}), and the intermediate line is for the $gg \to {\bar s}s$ case $f_{gg}(x)$ given by
(\protect\ref{ggf}).
\label{fig:f}}}
\end{figure}

Graphs of the correlation function $f(x)$ for the different production mechanisms considered are shown in Fig.~\ref{fig:f}.
We see that a clear distinction can be drawn between the scalar/pseudoscalar case, for which $f(x) = - 1$
for all $x$, and the vector case, for which $1 \ge f(x) \ge 1/3$. We return later to the $gg \to {\bar s}s$
case, which is shown as the intermediate line in Fig.~\ref{fig:f}.

\subsection{Production via $gg$ Fusion}

In this case there are three perturbative diagrams at lowest order: two `QED-like'
diagrams with $s$-quark exchanges in the $t$ and $u$ channels, and one
distinctively non-Abelian diagram with direct $s$-channel gluon exchange. By itself,
the latter would yield a vector coupling to ${\bar s}s$, akin to the previous ${\bar q} q$ case,
but other couplings are made possible by the other diagrams, and become important if
$m_s$ cannot be neglected. For example, if two gluons
with the same helicity collide with $\sqrt{s} = 2 m_s$, they may produce an ${\bar s}s$
via an effective scalar or pseudoscalar coupling.

Although it is a trivial standard calculation~\cite{MahlonParke}~\footnote{See~\cite{Combridge}
for the unpolarized case.}, for easy reference we include here the
full squared amplitude for $gg \to {\bar q}q$ where $q$ is a generic massive quark,
summed over final colours and averaged over initial colours and polarizations, in the form
\begin{equation}
\Sigma |{\cal M}|^2\ \; = \; \ {1\over 4}\ \pi^2 {\alpha}_S^2\ ({\cal F}+\lambda {\cal G})\ ,
\label{a1}
\end{equation}
where $\lambda = -1$ when the polarizations of the $q$ and ${\bar q}$ are the same,
$\lambda = +1$ when the polarizations of the $q$ and ${\bar q}$ are opposite, and
we work in the centre-of-mass frame of the $gg$ and ${\bar q}q$ pairs.
The coefficients ${\cal F}$ and ${\cal G}$ in (\ref{a1}) are given by
\begin{eqnarray}
{\cal F}&=&
+\ {8\over 3}\
{E^2(E^2-p^2{\rm cos}^2\theta ) +m^2E(E-p{\rm cos}\theta)-m^4\over
E^2(E-p{\rm cos}\theta)^2}
\nonumber\\
&&+\ {8\over 3}\
{E^2(E^2-p^2{\rm cos}^2\theta ) +m^2E(E+p{\rm cos}\theta)-m^4\over
E^2(E+p{\rm cos}\theta)^2}
\nonumber\\
&&+\ (-{1\over 8}) \times {16\over 3}\
{m^2p^2\over
E^2(E-p{\rm cos}\theta)(E+p{\rm cos}\theta)}
\nonumber\\
&&+\ 3\ {(E^2-p^2{\rm cos}^2\theta ) \over E^2}
\nonumber\\
&&-\ 3\ {E^2(E^2-p^2{\rm cos}^2\theta)-m^2Ep{\rm cos}\theta \over
E^3(E-p{\rm cos}\theta ) }
\nonumber\\
&&-\ 3\ {E^2(E^2-p^2{\rm cos}^2\theta)+m^2Ep{\rm cos}\theta \over
E^3(E+p{\rm cos}\theta ) }
\label{a2}
\end{eqnarray}
and
\begin{eqnarray}
{\cal G}&=&
+\ {8\over 3}\
{(-p^4+E^4{\rm cos}^2\theta -m^2Ep{\rm cos}\theta)\over
E^2(E-p{\rm cos}\theta)^2}
\nonumber\\
&&+\ {8\over 3}\
{(-p^4+E^4{\rm cos}^2\theta +m^2Ep{\rm cos}\theta)\over
E^2(E+p{\rm cos}\theta)^2}
\nonumber\\
&&+\ (-{1\over 8}) \times {16\over 3}\
{m^2 \Big( E^2(1-{\rm cos}^2\theta)+p^2 \Big) \over
E^2(E-p{\rm cos}\theta)(E+p{\rm cos}\theta)}
\nonumber\\
&&-\ 3\ {E^2(1-{\rm cos}^2\theta ) -m^2{\rm cos}^2\theta \over E^2}
\nonumber\\
&&+\ 3\ {E^4(1-{\rm cos}^2\theta)+m^2Ep{\rm cos}\theta -m^2E^2{\rm cos}^2\theta \over
E^3(E-p{\rm cos}\theta ) }
\nonumber\\
&&+\ 3\ {E^4(1-{\rm cos}^2\theta)-m^2Ep{\rm cos}\theta -m^2E^2{\rm cos}^2\theta \over
E^3(E+p{\rm cos}\theta ) } \, ,
\label{a3}
\end{eqnarray}
where $E$, $p$, $m$ are the energy, magnitude of 3-momentum and mass of the final-state
quark or antiquark, respectively, and $\theta$ is the angle between the 3-momentum of one of the initial gluons
and that of final-state quark.

The correlation function $f(x)$ defined in (\ref{dd7}) is given in this case by
\begin{equation}
f_{gg}(x) \; = \; \frac{({F}-{G})-({F}+{G})}{({F}-{G})+({F}+{G})} \;
= \; - \,\, \frac{G}{F} \, ,
\label{ggf}
\end{equation}
where $F=\int_{-1}^{+1}d({\rm cos}\theta){\cal F}$ and $G=\int_{-1}^{+1}d({\rm cos}\theta){\cal G}$ with
${\cal F}$ and ${\cal G}$ given in (\ref{a2}) and (\ref{a3}), respectively.
We note that ${\cal F}$ in (\ref{a2}) agrees with the formula presented in Ref. \cite{Combridge}
for the spin-summed squared amplitude.

The sum of the first three terms in (\ref{a2}) and (\ref{a3}) is proportional to the formula
for QED if we drop the relative color factor $(-{1\over 8})$ in their third terms.
The value of $f_{gg}(x) = - \, {G}/{F}$ for $gg \to {\bar s}s$ is shown in Fig.~\ref{fig:f} as a function of $x = 2 m_s/\sqrt{s}$.
As expected, we see that the vector case $f \to 1$ is recovered in the massless limit $x \to 0$, 
whereas the scalar/pseudoscalar
case $f \to - 1$ is recovered in the non-relativistic limit $x \to 1$, and $f_{gg}$ interpolates
monotonically between these limits for intermediate $x$~\footnote{Similar behaviour for ${\bar t} t$ production
has been emphasized and discussed in~\cite{MahlonParke}.}.

\section{Summary and Discussion}

We have pointed out that $\Lambda {\bar \Lambda}$ spin correlations offer, in principle,
an interesting window into the hadronization process, as possible fossils of the
spin correlations of their ancestral ${\bar s} s$ pairs. We have shown that
$\Lambda {\bar \Lambda}$ pairs produced via perturbative vector couplings to ${\bar s}s$ could have very
different spin correlations from those produced via non-perturbative scalar or pseudoscalar couplings
to ${\bar s}s$. The spin correlations of $\Lambda {\bar \Lambda}$ pairs produced perturbatively via
$gg$ collisions would be intermediate, tending towards the vector case if $m_s$ could be neglected,
and towards the scalar/pseudoscalar case in the limit of non-relativistic ${\bar s}s$ pairs.

A detailed discussion of the experimental possibilities for measuring these correlations lies
beyond the scope of this paper, but we emphasize that the ${\bar s}s$ production mechanisms
might be quite different in different kinematic regimes. For example, ${\bar s}s$ pairs produced
in high-$p_T$ jets might have a more `perturbative' origin, whereas those produced in minimum-bias
or heavy-ion collisions might have a more `non-perturbative' origin. It would therefore be interesting
to compare and contrast any $\Lambda {\bar \Lambda}$ spin correlations measured in these different
conditions.

Superficial consideration of the LHC experiments suggests that ALICE~\cite{ALICE} may be best suited for
measurements of $\Lambda {\bar \Lambda}$ spin correlations in minimum bias and low-$p_T$
heavy-ion collisions, whereas ATLAS~\cite{ATLAS} and CMS~\cite{CMS} may be better suited for measurements at higher $p_T$.
We emphasize that the $\Lambda {\bar \Lambda}$ pairs of interest are those with the lowest possible
invariant mass, which are most likely to originate from the same `parent' ${\bar s}s$ pair. Pairs with
larger relative momenta are not expected to exhibit any significant spin correlations.

\subsubsection*{Acknowledgements}

The work of J.E. is supported partly by the London
Centre for Terauniverse Studies (LCTS), using funding from the European
Research Council 
via the Advanced Investigator Grant 267352.
The work of D.S.H. is supported partly by
Korea Foundation for International Cooperation of Science \& Technology (KICOS)
and National Research Foundation of Korea (2011-0005226).
D.S.H. thanks CERN for its hospitality while working on this subject.
We thank Adam Jacholkowski for discussions of the experimental possibilities with ALICE,
and Homer Neal and Daniel Scheirich for discussions of the experimental possibilities with ATLAS.

\end{document}